\documentclass[showpacs,eqsecnum,aps,nofootinbib,floatfix]{revtex4}

\usepackage{citesort,subfigure,epsfig}

\def  \btosll      {$b \to s \ell^- \ell^+$}

\def  \btokll      {$B \to K \ell^- \ell^+$}
\def  \btoktt      {$B \to K \tau^- \tau^+$}

\def  \beq         {\begin{equation}}
\def  \eeq         {\end{equation}}
\def  \beqa        {\begin{eqnarray}}
\def  \eeqa        {\end{eqnarray}}
\def  \bcen        {\begin{center}}
\def  \ecen        {\end{center}}
\def  \bfig        {\begin{figure}}
\def  \efig        {\end{figure}}
\def  \btab        {\begin{tabular}}
\def  \etab        {\end{tabular}}

\def  \vtbvts      {V_{tb} V_{ts}^*}

\def  \sh          {\hat{s}}
\def  \mlh         {\hat{m}_{\ell}}
\def  \hatmb       {\hat{m}_b}
\def  \hatmk       {\hat{m}_{K}}

\def  \cseveff     {C_7^{eff}}
\def  \cneff       {C_9^{eff}}
\def  \cten        {C_{10}}
\def  \cqone       {C_{Q_1}}
\def  \cqtwo       {C_{Q_2}}

\def  \ev#1        {{\bf e}_#1}
\def  \wv#1        {{\bf w}_#1}
\def  \pmv         {{\bf p_-}}
\def  \ppv         {{\bf p_+}}
\def  \pkv          {{\bf p_{K}}}

\def  \etal        {{\it et al.}}

\def \prd#1#2#3       {Phys. \ Rev. {\bf D #1}, {#2} (#3)}
\def \prl#1#2#3       {Phys. \ Rev. \ Lett. {\bf #1}, {#2} (#3)}
\def \nuclphysb#1#2#3 {Nucl. \ Phys. {\bf B #1}, {#2} (#3)}
\def \plb#1#2#3       {Phys. \ Lett. {\bf B #1}, {#2} (#3)}
\def \physrep#1#2#3   {Phys. \ Rep {\bf #1}, {#2} (#3)}
\def \zphysc#1#2#3    {Z. \ Phys. {\bf C #1}, {#2} (#3)}

\begin{document}




\begin{flushright}
\begin{tabular}{l}
hep-ph/0307276, \\
KIAS-P03057   \\
\end{tabular}
\end{flushright}



\title{Supersymmetric effects on the Forward Backward asymmetries of $B
\to K \tau^+ \tau^-$}

\author{S. Rai Choudhury}
  \email{src@physics.du.ac.in}
   \affiliation{Department of Physics \& Astrophysics \\
        University of Delhi, Delhi - 110 007, India}
\author{A. S. Cornell}
  \email{alanc@kias.re.kr}
   \affiliation{Korea Institute of Advanced Study, Cheongryangri 2-dong, \\
      Dongdaemun-gu, Seoul 130-722, Republic of Korea}
\author{Naveen Gaur}
  \email{naveen@physics.du.ac.in}
   \affiliation{Department of Physics \& Astrophysics \\
        University of Delhi, Delhi - 110 007, India}
\author{G. C. Joshi}
  \email{joshi@tauon.ph.unimelb.edu.au}
   \affiliation{School of Physics, University of Melbourne, \\
        Victoria 3010, Australia}



\begin{abstract}
Leptonic and semi-leptonic rare decays of B-mesons provide
significant (both theoretically and experimentally) signatures of
any new physics beyond the Standard Model (SM).  More specifically
the decay \btokll has been theoretically observed to be very
sensitive to new physics, as the Forward Backward (FB)
asymmetry in this decay mode vanishes in the SM.  Supersymmetry,
however, predicts a non-vanishing value of this asymmetry.  In
this work we will study the polarized lepton pair FB asymmetry,
i.e. the FB asymmetry of the lepton when one (or both) final state
lepton(s) are polarized. We will study these asymmetries both
within the SM and for Supersymmetric corrections to the SM.
\end{abstract}

\maketitle


\setcounter{footnote}{0}

\section{\label{section:1} Introduction}

Lately there has been enormous progress in the study of flavour
physics, where the B system has provided us with one of the most
ideal environments for this type of study. Of the decay modes
considered, theoretically and experimentally, the study of the
{\em ``rare decays''} of the B-mesons are of particular
interest. Here the name {\em rare} has been given to those decay
modes which arise from Flavour Changing Neutral Currents (FCNC).
FCNC processes are absent at the tree level in the Standard Model
(SM) but can occur through loop diagrams, their strength
being proportional to the Fermi Constant. FCNC processes
involving, for example $b \to s$ and $b \to d$
transitions are therefore more sensitive to the
details of the SM interactions and thus are suited to the study of
possible new physics beyond the SM. Theoretically inclusive FCNC
processes like $B \to X_{s(d)} \ell^+ \ell^-$ are relatively
cleaner than their exclusive counterparts, since they are
relatively independent of the quark structure of the hadrons
involved. They are however difficult to measure experimentally
(for details refer to Chapter 7 in reference \cite{BTeV:2002a})
and one may expect a substantial amount of experimental
information regarding various exclusive B decay processes from the
B-factories. Amongst the important FCNC exclusive B-decay
processes are $B \rightarrow charmless$ $meson + \gamma$ and $B
\rightarrow charmless$ $meson + lepton$ $pair$. This second
process potentially provides a very rich set of
experimental observables involving various momenta and spin
polarization correlations. It is thus important to theoretically
calculate all possible measurable parameters of these processes.

\par In recent times there have been many calculations of such
processes like $B\to K(K^*) \ell^+ \ell-$
\cite{Geng:pu,Aliev:2001pq,Aliev:2002ux}, $B \to \pi (\rho) \ell^+
\ell^-$ \cite{Iltan:1998a}, $B_{s,d} \to \ell^+ \ell^-$
\cite{Choudhury:1999ze} and $B_{s,d} \to \ell^+ \ell^-
\gamma$ \cite{RaiChoudhury:2002hf}. Amongst these the ones
involving  a quark level $b \to s$ transition are expected to have
relatively large branching ratios. For the $B \to K^* \ell^+
\ell^-$ transition possible experimentally accessible parameters
like Forward-Backward asymmetry, lepton polarization asymmetry
etc., have been studied
\cite{Ali:1999mm,Kruger:1996cv,Geng:pu,Aliev:2002ux}. In
particular, polarization correlations between the two leptons,
which was suggested recently by Bensalem \etal
\cite{Bensalem:2002ni} have also been studied in the context of
this process \cite{Choudhury:2003a}. In this note we carry out an
analysis of the polarized Forward Backward (FB) asymmetries in the
process $B \to K \ell^+ \ell^-$.

\par The theoretical basis for the study of FCNC B-decay processes is
now a well established formalism based on the operator
product expansion and use of the renormalization group
\cite{Buchalla:1995vs}. The formalism ultimately produces an
effective Hamiltonian for every process involving low
dimensional hadronic operators, in the form of
currents, with numerical multiplying coefficients called the
Wilson coefficients. The Wilson coefficients encrypt short
distance properties of the weak Hamiltonian and are sensitive to
physics beyond the SM at high energy scales, in particular, to
supersymmetric extensions of the SM. A great deal of
theoretical work has gone, in recent times, to
evaluating their values both within the SM and in the context of
the minimal extension of the standard model (MSSM). Evaluation of
the matrix elements of the hadronic currents on the other hand
involve the relatively long distance quark structure of the
hadrons. The most ideal way to evaluate them would be through
lattice gauge theory calculations, which do not exist for all
B-meson processes at the moment. Alternatively, one relies on
evaluations based on Light cone sum rules and these also have been
compiled for a variety of processes. We shall make extensive use
of these in our calculations.

\par This paper is organized as follows:  In section \ref{section:2}
we will discuss the effective Hamiltonian for the process under
consideration. In section \ref{section:3} we will introduce our
notation for the polarized FB asymmetry.  Finally, in
section \ref{section:4} we will present results of our numerical
analysis and the conclusions.

\section{\label{section:2} Effective Hamiltonian}

In this paper we are interested in the process \btokll , which has the
basic quark level transition $b \to s$. The  
effective Hamiltonian for a such transition has been summarized in
the literature \cite{Buchalla:1995vs}. The effective Hamiltonian
is arrived at by integrating out the heavy degrees of freedom from
the full theory. In the SM the heavy degrees of freedom are
$W^\pm, Z$ and the top quark; in the MSSM all the new SUSY
particles shall also be counted. From such considerations we
arrive at the effective Hamiltonian \cite{Xiong:2001up} : 
\beq
{\cal H}_{eff} = \frac{4 G_F}{\sqrt{2}} \vtbvts
         \Bigg[ \sum_{i = 1}^{10} C_i(\mu) O_i(\mu)
         + \sum_{i = 1}^{10} C_{Q_i}(\mu) Q_i(\mu)
         \Bigg]
\label{sec2:eq:1} 
\eeq 
where the $O_i$ are the current-current ($i
= 1,2$), penguin ($i = \mathbf{3},\dots,6$), magnetic penguin ($i
= 7,8$) and semi-leptonic ($i = 9,10$) operators, whereas
the $C_i(\mu)$ are the corresponding Wilson coefficients
renormalized at scale $\mu$. The value of these coefficients have been
given in references \cite{Grinstein:1989me,Cho:1996we}.  The
additional operators $Q_i$ $(i = 1,\dots 10)$, and their corresponding
Wilson coefficients are due to the Neutral Higgs boson (NHB) exchange 
diagrams and are given in references
\cite{Choudhury:1999ze,Xiong:2001up}.

\par Different FCNC decays involve different combinations of $C_i$'s
and $C_{Q_i}$'s and thus provide us with independent information on
these coefficients.  At the quark level the transition $b \to s
\gamma$ is sensitive only to the magnitude of $C_7$, whereas the
semi-leptonic transition \btosll is sensitive to $C_9, C_{10},
C_{Q_1}$ and $C_{Q_2}$ as well\footnote{It has also been shown in
many works that \btosll is sensitive even to the signs of these
Wilson coefficients and hence this decay channel will provide
information not only on the magnitude but also the sign of these
coefficients.}. 

\par From the effective Hamiltonian given in equation
(\ref{sec2:eq:1}) the decay amplitude for \btokll is calculated to be:
 \beqa
{\cal M} = && \frac{\alpha G_F}{\sqrt{2} \pi}
\vtbvts
  \left\{
     - 2 \cseveff \frac{m_b}{q^2} (\bar{s} i \sigma_{\mu \nu}
      q^\nu P_R b) (\bar{\ell} \gamma^\mu \ell)
     + \cneff (\bar{s} \gamma_\mu P_L b) (\bar{\ell} \gamma^\mu \ell)
     + \cten (\bar{s} \gamma_\mu P_L b) (\bar{\ell} \gamma^\mu
        \gamma_5 \ell) \right.            \nonumber   \\
   &&  \left.
     + \cqone (\bar{s} P_R b) (\bar{\ell} \ell)
     + \cqtwo (\bar{s} P_R b) (\bar{\ell} \gamma_5 \ell)
   \right\}
\label{sec2:eq:2} \eeqa where $q$ is the momentum transferred to
the lepton pair, given as $q = p_- + p_+$, where $p_-$ and
$p_+$ are the momentas of $\ell^-$ and $\ell^+$ respectively. The
$\vtbvts$ are the CKM factors and $P_{L,R} = (1 \mp \gamma_5)/2$.
In writing the above matrix element (and in future analysis) we
will neglect the mass of the strange quark, whereas lepton masses
shall be retained.

\par The free quark decay amplitude given in equation (\ref{sec2:eq:2})
contains certain long distance effects which are absorbed in the
redefinition of the $C_9$ Wilson coefficient (where we use the
prescription given in reference \cite{Long-Distance}) : 
\beq 
\cneff (\sh) = C_9 + Y(\sh) . 
\label{sec2:eq:3} 
\eeq 
The $Y(\sh)$ part has a perturbative as well as a non-perturbative
part. The origin of the non-perturbative part is from the
resonance corrections to the perturbative quark loops (which gives
the perturbative contribution to $Y(\sh)$).  We will also use the
usual Breit-Wigner prescription to take care of the resonant
contribution \cite{Kruger:1996cv,Long-Distance}.  This
prescription implies adding  resonant terms  to $\cneff$: 
\beq
C_9^{res} \propto \kappa \sum_{V = \psi}
         \frac{\hat{m}_V Br(V \to \ell^- \ell^+)
       \hat{\Gamma}^V_{total}}{\sh - \hat{m}_V^2 + i \hat{m}_V
        \hat{\Gamma}^V_{total}}
\label{sec2:eq:4} 
\eeq 
where all the symbols above have been explained in the work of
Kr\"{u}ger and Sehgal \cite{Kruger:1996cv}. For the phenomenological
factor, $\kappa$, we will choose a value 2.3.

\par Using the definitions of the form factors given in Appendix
\ref{appendix:a} we can write the matrix element given in equation
(\ref{sec2:eq:2}) as: \beq {\cal M} = \frac{\alpha 
G_F}{2 \sqrt{2} \pi} \vtbvts \Bigg\{ A (p_K)_\mu (\bar{\ell}
\gamma^\mu \ell) + B (p_K)_\mu (\bar{\ell} \gamma^\mu \gamma_5
\ell) + C (\bar{\ell} \ell) + D (\bar{\ell} \gamma_5 \ell) \Bigg\}
\label{sec2:eq:5a} \eeq where the coefficients $A, B, C$ and $D$
in equation (\ref{sec2:eq:6}) are given as\footnote{In writing
equation (\ref{sec2:eq:6}) we have used the equations of motion:
$p_\mu \bar{\ell} \gamma^\mu \ell = 0$; $p_\mu \bar{\ell}
\gamma^\mu \gamma_5 \ell = 2 m_\ell (\bar{\ell} \ell)$.}: 
\beqa 
A
&=&  4 f_T \frac{\hatmb \cseveff}{1 + \hatmk} + 2
 f_+ \cneff
\label{sec2:eq:7}  \\
B &=&  2 f_+ C_{10}
\label{sec2:eq:8} \\
C &=&  f_0 \frac{1 - \hatmk^2}{\hatmb} \cqone
\label{sec2:eq:9} \\
D &=& f_0 \frac{1 - \hatmk^2}{\hatmb} \cqtwo + 2 \mlh C_{10} f_+ -
2 \mlh f_+ \frac{1 - \hatmk^2}{\sh} C_{10} + 2 \mlh f_0 \frac{1 -
\hatmk^2}{\sh} C_{10}
\label{sec2:eq:10}
\eeqa

\par With the above expression of the matrix element, equation
(\ref{sec2:eq:5a}), we can obtain the expression for
the differential decay rate as: \beq \frac{d \Gamma}{d \sh} =
  \frac{\alpha^2 G_F^2 m_B^5}{2^{12} \pi^5} |\vtbvts|^2 \lambda^{1/2}
\sqrt{1 - \frac{4 \mlh^2}{\sh}} ~ \bigtriangleup
\label{sec2:eq:5}
\eeq
where
\beqa
\bigtriangleup
&=&
{2 \over 3} \lambda \left(1 + \frac{2 \mlh^2}{\sh} \right) |A|^2
+ {2 \over 3} \Bigg[ \lambda \left(1 + \frac{2 \mlh^2}{\sh} \right)
 + 24 \hatmk^2 \mlh^2 \Bigg] |B|^2 + 4 (\sh - 4 \mlh^2) |C|^2
+ 4 \sh |D|^2         \nonumber \\
&& + 8 \mlh (1 - \hatmk^2 - \sh) Re(B^* D) \label{sec2:eq:6} \eeqa
with $\lambda = 1 + \hatmk^4 + \sh^2 - 2 \hatmk^2 \sh - 2 \hatmk^2
- 2 \sh$. In the next section we will use this expression
for the differential decay rate to introduce and
define the polarized FB asymmetries, followed by our analytical
expressions for these asymmetries.


\section{\label{section:3} Polarized FB asymmetries}

Firstly we will define the polarization vectors of $\ell^-$ and
$\ell^+$, where in this definition we will use the
convention followed in many earlier works
\cite{Kruger:1996cv,Choudhury:2003a,Geng:pu,Aliev:2001pq}. In
order to evaluate the polarized FB asymmetries we introduce a spin
projection operator defined by $N = (1 + \gamma_5 {\not S}_x)/2$
for $\ell^-$ and $M = (1 + \gamma_5 {\not W}_x)/2$ for $\ell^+$,
where $x = L$, $N$, or $T$ (corresponding to the
longitudinal, normal and transverse polarization asymmetries
respectively ). Next we define the orthogonal unit
vectors $S_x$ for $\ell^-$ and $W_x$ for $\ell^+$ in the rest
frames of $\ell^-$ and $\ell^+$ respectively as:  
\beqa 
S^\mu_L
&\equiv& (0, \ev{L}) ~=~ \left(0, \frac{\pmv}{|\pmv|} \right)
                 \nonumber               \\
S^\mu_N &\equiv& (0, \ev{N}) ~=~ \left(0, \frac{\pkv \times \pmv}{|\pkv
             \times \pmv |}\right)
                 \nonumber               \\
S^\mu_T &\equiv& (0, \ev{T}) ~=~ \left(0, \ev{N} \times \ev{L}\right)
            \label{sec3:eq:1a}             \\
W^\mu_L &\equiv& (0, \wv{L}) ~=~ \left(0, \frac{\ppv}{|\ppv|}\right)
                 \nonumber                \\
W^\mu_N &\equiv& (0, \wv{N}) ~=~ \left(0, \frac{\pkv \times
                \ppv}{|\pkv \times \ppv |} \right)
                 \nonumber                \\
W^\mu_T &\equiv& (0, \wv{T}) ~=~ (0, \wv{N} \times \wv{L})
                 \label{sec3:eq:2a}
\eeqa 
where $\pmv, \ppv$ and $\pkv$ are the three momentas of
$\ell^-, \ell^+$ and the K-meson in the dilepton CM frame.  From
the rest frames of the leptons we boost the four vectors $S_x$ and
$W_x$ to the dilepton CM frame.  Only the longitudinal vectors,
$S_L$ and $W_L$ will be boosted by the Lorentz transformation, to a
value of : 
\beqa
 S^\mu_L &=& \left( 
\frac{|\pmv|}{m_\ell}, \frac{E_\ell \pmv}{m_\ell
                  |\pmv|}   \right)
           \nonumber               \\
W^\mu_L &=& \left( \frac{|\pmv|}{m_\ell}, - \frac{E_\ell \pmv}{m_\ell
                |\pmv|} \right)
\label{sec3:eq:3a} 
\eeqa 
where $E_\ell$ is the energy of either of the leptons (both having the
same energy in this frame) in the dileptonic CM frame.

\par The definition of the differential Forward-Backward (FB)
asymmetry is given in references \cite{Ali:1991is,Ali:1999mm}:
\beq
\overline{A}(\sh) = \int_0^1 \frac{d^2
\Gamma}{d\sh dz} dz   - \int_{-1}^0 \frac{d^2 \Gamma}{d\sh dz} dz
.
\label{sec3:eq:1}
\eeq

\noindent Consider the case where we shall not sum over the spins
of the outgoing leptons.  In general the FB asymmetry will be a
function of the spins of the final state leptons, and as such can
be defined as :
 \beq 
\overline{A}(s^-, s^+, \sh) = \int_0^1
\frac{d^2 \Gamma(s^-, s^+)}{d\sh dz} dz - \int_{-1}^0 \frac{d^2
\Gamma(s^-, s^+)}{d\sh dz} dz . 
\label{sec3:eq:2} 
\eeq 
From an experimental viewpoint the normalized FB asymmetry is more
useful. Therefore we shall normalize the above expression (equation
\ref{sec3:eq:2}) for the FB asymmetry 
by dividing by the total decay rate: 
\beq A(s^-, s^+, \sh) =
\frac{\overline{A}(s^-, s^+ ,\sh)}{d\Gamma/d\sh} .
\label{sec3:eq:3} 
\eeq 
In analogy to the prescription given in
Bensalem \etal \cite{Bensalem:2002ni} we can split this FB
asymmetry into various polarization components\footnote{The
convention followed is that the repeated index is summed over.}:
\beq A(s^-,s^+) =
 A + A_i^- s_i^- + A_i^+ s_i^+ +
A_{ij} s_i^+ s_j^-
\label{sec3:eq:4}
\eeq
\noindent where $i,j = L,T,N $ are the longitudinal, transverse and
normal components of the polarization.  Using this definition we can
write the single and double lepton polarized FB asymmetries.  From
equation (\ref{sec3:eq:4}) the single polarized lepton FB
asymmetry can be written as:
\beqa
A_i^- = A(s^- = i, s^+ = j) +
A(s^- = i , s^+ = - j) - A(s^- = - i, s^+ = j)
- A(s^- = - i , s^+ = - j)
\label{sec3:eq:5}   \\
A_i^+ = A(s^- = j, s^+ = i) + A(s^- = - j , s^+ = i) - A(s^- = j,
s^+ = - i) - A(s^- = - j , s^+ = - i) \label{sec3:eq:6} \eeqa
\noindent and the doubly polarized FB asymmetry can be written as:
\beq A_{ij} = A(s^- = i, s^+ = j) - A(s^- = i , s^+ = -j ) - A(s^-
= -i, s^+ = j) + A(s^- = - i , s^+ = - j) . \label{sec3:eq:7} \eeq
Using the above expressions of the FB asymmetries the results of the
unpolarized FB asymmetry is evaluated to be :   
\beq A = 2
\mlh \lambda \sqrt{1 - \frac{4 \mlh^2}{\sh}} \frac{Re(A^*
C)}{\bigtriangleup} . 
\label{sec3:eq:8} 
\eeq
\noindent From the  expression given above and equation
(\ref{sec2:eq:9}), we can see that the unpolarized FB asymmetry is
proportional to $\cqone$.  This point has been emphasized in many 
earlier works \cite{Bobeth:2001sq,Iltan:1998a}. In 
the SM $\cqone$ is absent and hence for the decay modes $B \to K (\pi)
\ell^+ \ell^-$ the FB asymmetry within the SM vanishes. However, in
SUSY (and 2HDM) extensions of the SM there exists a non-vanishing
value of $\cqone$ , and hence a non-vanishing value of the FB 
asymmetry \cite{Bobeth:2001sq,Iltan:1998a} is possible. Therefore a 
non-vanishing value of the FB asymmetry can be regarded as clear
signal of new physics beyond the SM.

\par The analytical results of the polarized FB asymmetries are :
\beqa
A^-_L
&=&
\frac{4 \lambda^{1/2} \mlh}{\bigtriangleup}
 \Bigg[ \mlh (-1 + \hatmk^2 + \sh) Re(A^* B) - \mlh Re(A^* D) \Bigg]
  \label{sec3:eq:9} \\
A^-_N &=& 0
  \label{sec3:eq:10} \\
A^-_T &=&
 \frac{4 \mlh}{3 \sh \bigtriangleup}
 \sqrt{1 - \frac{4 \mlh^2}{\sh}} \lambda Re(A^* B)
  \label{sec3:eq:11} \\
A^+_L &=&  A^-_L
  \label{sec3:eq:12} \\
A^+_N &=& 0
  \label{sec3:eq:13} \\
A^+_T &=& A^-_T
  \label{sec3:eq:14} \\
A_{LL} &=&  A
  \label{sec3:eq:15} \\
A_{LN} &=&
  \frac{4 \mlh}{3 \sqrt{\sh} \bigtriangleup} \sqrt{1 - \frac{4
   \mlh^2}{\sh}} \lambda  Im(A^* B)
  \label{sec3:eq:16}  \\
A_{LT} &=&
   \frac{4 \mlh}{3 \sqrt{\sh}}
   \sqrt{1 - \frac{4 \mlh^2}{\sh}} \lambda \frac{|A|^2}{\bigtriangleup}
  \label{sec3:eq:17} \\
A_{NL} &=& A_{LN}
  \label{sec3:eq:18} \\
A_{NN} &=& - A
  \label{sec3:eq:19} \\
A_{NT} &=&
  4 \mlh \lambda^{1/2}
\frac{ (1 - \hatmk^2 - \sh) \mlh Im(A^* B) + Im(A^*
D)}{\bigtriangleup}
  \label{sec3:eq:20} \\
A_{TL} &=& - A_{LT}
  \label{sec3:eq:21} \\
A_{TN} &=& A_{NT}
  \label{sec3:eq:22} \\
A_{TT} &=&  A
  \label{sec3:eq:23}
\eeqa
where $\bigtriangleup$ is given in equation (\ref{sec2:eq:6}) and
$A$ is the unpolarized FB asymmetry given in equation
(\ref{sec3:eq:8}).  We will discuss the above obtained expressions
of the various FB asymmetries and present our numerical analysis
of the same in the next section.


\section{\label{section:4} Numerical analysis, results and conclusion}

In this section we shall present our numerical analysis of the
observables whose analytical expressions were given in the
previous section. We will also present the variation of all
the observables with the dilepton invariant mass.

\par As it is experimentally more useful to have the average values
of these quantities we shall present our results as the
averages values of these quantities, where we will define
our averages as:
\begin{equation}
\langle
A \rangle \equiv \frac{\displaystyle{\int_{(3.646 +
0.02)^2/m_B^2}^{(m_B - m_{K})^2/m_B^2}} A \frac{d
\Gamma}{d \sh} d \sh}{ \displaystyle{\int_{(3.646 +
0.02)^2/m_B^2}^{(m_B - m_{K})^2/m_B^2}} \frac{d \Gamma}{d \sh} d
\sh}
\label{sec4:eq:1}
\end{equation}
which means that in calculating the average we have taken
the lower limit of integration to be above the first
resonance\footnote{The first resonance here means the
resonance after the threshold of the decay, which is $s \ge 4
m_\tau^2$.}. The input parameters of our numerical analysis are
defined in Appendix \ref{appendix:b}, and our SM predictions of
the integrated observables are given in Table \ref{sec4:tab:1}.
\begin{table}
\begin{center}
\begin{tabular}{| c | c | c | c | c | c | c |}
\hline
\hspace{.3cm} Br(\btoktt) \hspace{.3cm} &
\hspace{.3cm} $A$ \hspace{.3cm} &
\hspace{.3cm} $A^-_L$ \hspace{.3cm} &
\hspace{.3cm} $A^-_T$ \hspace{.3cm} &
\hspace{.3cm} $A_{LN}$ \hspace{.3cm} &
\hspace{.3cm} $A_{LT}$ \hspace{.3cm} &
\hspace{.3cm} $A_{NT}$ \hspace{.3cm}  \\
\hline
$1.17 \times 10^{-7}$  &
0 & 0.363 & -0.097 & 0.023 & 0.187 & .0847 \\
\hline
\end{tabular}
\caption{The SM predictions for the integrated observables}
\label{sec4:tab:1}
\end{center}
\end{table}

\par Before discussing our results we shall first elaborate on
the models in which we have performed our numerical analysis.  We
have worked with the minimal supersymmetric extension of the
standard model (MSSM), this being the simplest SUSY extension of
the SM with the least number of parameters introduced
\cite{Nilles:1983ge}. But even in the MSSM we are required to
introduce a large number of parameters, over and above the number
of parameters in the SM. To ease out this problem and to reduce
such a large number of parameters to a more
manageable level, many models have been introduced such as the
dilaton, moduli, minimal Supergravity (mSUGRA) 
\cite{minsugra}, AMSB (Anomaly mediated SUSY breaking)
\cite{amsbmod} and the GMSB (Gauge mediated SUSY breaking)
\cite{gmsbmod} models. The generic feature of all these models
 is that they assume some sort of unification of the
parameters of the MSSM at a higher scale. In the
literature the mSUGRA model is also known as the CMSSM
(Constraint MSSM) \cite{cmssmmod}. We shall further assume
that the soft SUSY breaking parameters are real. For our
numerical analysis we will use two types of SUGRA (Supergravity)
models, namely mSUGRA and rSUGRA (relaxed SUGRA) which we describe. In
both these models it is believed that the SUSY breaking occurs in a
hidden sector and is communicated to the visible sector only by  
gravitational-strength interactions. As such, soft breaking
terms are assumed to be flavour blind (like gravitational
interactions). 

\par In the mSUGRA model along with the unification of the coupling
constants $g_{1,2,3}$ (of the U(1), SU(2) and SU(3) gauge theories),
the other unification conditions are; 
\begin{itemize}
\item{} unification of the gaugino masses ($m_{1/2}$),
\item{} unification of the scalar (sfermion and Higgs) masses ($m_0$),
\item{} unification of the trilinear couplings (A),
\end{itemize}
all at the GUT scale. There are two other parameters. The first
is the ratio of vev (vacuum expectation value of two Higgs),
namely $tan\beta$. The second arises in the process of evolving the
soft SUSY breaking parameters 
from the GUT scale to the electroweak scale and
then imposing the correct low energy electroweak symmetry
breaking condition. This condition fixes the magnitude of
$\mu$ (the two Higgs coupling parameter), however, the sign
still remains uncertain \footnote{There are many conventions
followed regarding the sign of $\mu$, our convention is such that
$\mu$ appears in the chargino mass matrix with a positive sign
(where the $(g - 2)_\mu$ giving the parameter $\mu$ a negative sign is
disfavored.}. The sign of $\mu$ thus also enters taken as a
parameter. 

\par Therefore, in all mSUGRA frameworks we have five parameters,
namely:  
\beq 
m_{1/2} ~,~ m_0 ~,~ A ~,~ tan\beta ~~{\rm and}~~ sgn(\mu)
\nonumber 
\eeq 
Unification of all the scalars and also all the
gauginos is not an essential requirement of SUGRA models.  One can
have models where either all the scalars do not have a universal
mass at GUT scale or there is a non-universality of gaugino masses
at the GUT scale.  We shall also explore such a model, where
we would relax the condition of the universality of the scalar
masses at the GUT scale \cite{goto1}. In the literature these models
are known as non minimal SUGRA models \cite{goto1}. 
We will call such a model to be relaxed SUGRA
(rSUGRA) model. We will assume that the values of squarks and
the Higgs sector scalars masses different at the GUT scale.
This shall introduce another parameter into the model. This additional
parameter we will take as the mass of the pseudo-scalar Higgs
boson ($m_A$). However for our numerical analysis we will consider
only the region of the SUGRA parameter space which is consistent
with the $B \to X_s \gamma$ ~ 95\% CL
\cite{BTeV:2002a,Coan:2000a}:
$$
2 \times 10^{-4} \le Br(B \to X_s \gamma) \le 4.5 \times 10^{-4} .
$$
We present our results for the various decay rates and asymmetry
parameters considered in Figures (1)-(21). The branching fractions
of the decays considered are of course too low to be observed with
the current luminosities of the B-factories, but they will certainly
be possible in the foreseeable future.  As 
can be seen from the expressions of the polarized FB asymmetry,
they are sensitive to the Wilson coefficients that arise only
beyond the standard model, and thus a measurement of these
would be one more test of physics beyond the SM.

\par We now turn to the uncertainties in the estimates
 we obtained, where the Wilson coefficients we used are in the NLL
approximation and do not introduce 
significant uncertainties. The CKM parameters typically have
uncertainties of the order of 10\% \cite{Hagiwara:fs} while the
other SM parameters involved do not suffer from any large
uncertainties. The SUSY parameters are input parameters. For a
particular choice of parameters the major uncertainty in our
results arises is from the definition of the form factors used.
They typically have a 15\% uncertainty \cite{Beneke:2003zv}. There
are regions in the SUSY parameter space where SUSY effects
exceed the kind of errors which are introduced by the form
factor definition, for example the graph of the
branching ratio of the mode concerned \btokll in figures
\ref{fig:1}, \ref{fig:2}, \ref{fig:3}. Of course, at the present
level an uncertainty of ~ 20 \% modifications arising from SUSY
or any other physics beyond the SM, or levels 
less than this, would not be distinguishable from SM
results. We, however, emphasize that the null
polarization results which we have pointed out for some of the
double polarization asymmetries (for a mixture of $B_0$ and
$\bar{B_0}$) do not depend on the parameters. The experimental
deviation of these from null values would indicate a presence of
interactions beyond the one considered here, or may indicate
the source of CP-violation in the SUSY extension, for
example a complex value of $\mu$.

\par Regarding our numerical analysis we have presented
all the observables in section \ref{section:3}, where the
plots are given for all the possible observables with the
dilepton invariant mass ($\sh$), in figures \ref{fig:1} ,
\ref{fig:4} , \ref{fig:7} , \ref{fig:10}, \ref{fig:13} ,
\ref{fig:16} , \ref{fig:19} for the SM, mSUGRA and rSUGRA
models. We have also tested the sensitivity of the observables to
various MSSM parameters. For this purpose we have also presented
the results of the averaged values of the observables. For
averaging we have used the procedure defined in equation
(\ref{sec4:eq:1}). The plots of averaged observables with
$tan\beta$ for various values of $m_0$ are given in figures
\ref{fig:2}, \ref{fig:5}, \ref{fig:8}, \ref{fig:11}, \ref{fig:14},
\ref{fig:17}, \ref{fig:20} in the mSUGRA model. The other
model parameters are given in respective figure captions. In the
rSUGRA model we have plotted the averaged value of the 
observables as a function of pseudo-scalar Higgs mass ($m_A$) for
various values of $tan\beta$ in figures \ref{fig:3}, \ref{fig:6},
\ref{fig:9}, \ref{fig:12}, \ref{fig:15}, \ref{fig:18},
\ref{fig:21}.

\par As can be seen from the plots all the observables are very
sensitive to the various MSSM parameters, both in the mSUGRA and
rSUGRA models, as may have been expected. However, these plots of the
deviation of the observables from their respective SM values is more
pronounced for the rSUGRA model than the mSUGRA model. This is
essentially due to the additional parameter in the
rSUGRA model, which is controlled by the
pseudo-scalar Higgs mass. The value of the new Wilson coefficients
corresponding to scalar and pseudo-scalar operators is directly
proportional to $tan^3\beta$ and $m_A^2$ and lower the
value of $m_A$ the higher the value of those Wilsons.
Similarly, the higher the $tan\beta$ the higher the
new Wilsons. We effectively have that for
the rSUGRA  model a low $m_A$ and high $tan\beta$ region of MSSM
parameter space also available which can generate large values
of the new Wilsons. This was not the case in the mSUGRA model.

\par In figure \ref{fig:2} we have plotted of the integrated
branching ratio in the mSUGRA. The integrated branching
ratio can, at best, be 4 to 5 times the SM branching
ratio for low values of $m_{1/2}$ and very high values of
$tan\beta$. However, for the rSUGRA model there can be
a much higher enhancement to the branching ratio (as
given in figure \ref{fig:3}) when compared to the SM
value for low $m_A$ and high $tan\beta$.  As can be seen from the
other graphs of the averaged observables the 
various observables show marked deviation from their respective SM
values for a very wide region of the MSSM parameter space.
This deviation cannot be explained solely on the basis
of uncertainties in the form factors (which are at
worst 15\%). Thus such variations, if observed in future
B-factories, could be very useful in testing the underlying
operator structure of the theory and in fixing the numerical
value of the Wilson coefficients.

\par There is also a further aspect to our results in
relation to CP asymmetry. Consider the FB asymmetry of the
conjugate process  $b \to s \tau^+ \tau^-$, where due to the 
smallness of the coupling of the $b$-quark with the
$u$-quark the CKM-factor in all amplitudes
involving the $b \to s$ transition, like the present one, will
essentially be an overall factor. In the version of
supersymmetry that we have considered and the parameter space
thereof, there are no extra CP-violating phases. Thus in
calculating decay rates the phase will be washed away and we have
in effect a CP-invariant theory. The asymmetries of the process $b
\to s \tau^- \tau^+$ and the conjugate process $\bar b \to \bar s
\tau^+  \tau^- $ are thus related. In fact the unpolarized FB
asymmetry for the $b \to s$ transition will vanish in an
untagged (CP even) sample \cite{Choudhury:1997xa}. Defining the
forward and backward directions as referring to the $\tau^-$
and denoting the asymmetries of the conjugate process by
$\bar A $, we get: \beq \bar A_{ij} = p_{ij} A_{ji}
\label{sec4:eq:2} \eeq with the parity factor $p_{ij}$ equaling -1
for all $i's$ and $j's$ except for the combinations $LN$ and $NT$.
If we have an untagged sample containing an equal number of $B$'s
and $\bar{B}'s$, then just as in the unpolarized sample, the
asymmetries observed for the combinations $LL$, $NN$, $TT$ as well
as for $(LT+TL)$ will vanish. However, for the combinations
$(LN+NL)$ and $(NT+TN)$, the asymmetries will add up. We thus have
a situation in these two cases wherein a measurement of FB
asymmetry for an untagged sample can lead to a meaningful non-null
valued comparison between theory and experiment.


\appendix

\section{\label{appendix:a}Form Factors}

\setcounter{equation}{0}

\hspace{0.5cm} The form factors for the $B \to K$ transition are
given in reference \cite{Ali:1999mm}: \beqa \langle K(p_K) |
\bar{s} \gamma_\mu b | B(p_B) \rangle &=&  f_+ \Bigg[ (p_B +
p_K)_\mu - \frac{m_B^2 - m_K^2}{q^2} q_\mu \Bigg]
 + f_0 \frac{m_B^2 - m_K^2}{q^2} q_\mu
\label{appendixa:1}      \\
\langle K(p_K) | \bar{s} \sigma_{\mu \nu} q^\nu b | B(p_B)
\rangle &=& i \frac{f_T}{m_B + m_K} \Bigg[ (p_B + p_K)_\mu q^2 -
q_\mu (m_B^2 - m_K^2) \Bigg]
\label{appendixa:2}
\eeqa
where $q (=
p_+ + p_-)$ is the sum of four momentas $\ell^-$ and $\ell^+$,
i.e. the momentum transfer; $f_+$, $f_0$ and $f_T$ are the form
factors.  Multiplying equation (\ref{appendixa:1}) by $q^\mu$ and
by using the equations of motion we get: \beq \langle K(p_K) |
\bar{s} b | B(p_B) \rangle ~=~ f_0 \frac{(m_B^2 - m_K^2)}{m_b}
\label{appendixa:3} \eeq where all the other matrix elements
vanish.

\par For the form factors, $f_+, f_0$ and $f_T$ we will take the
parameterization: \beq F(\sh) = F(0) exp(c_1 \sh + c_2 \sh^2 + c_3
\sh^3) \label{appendixa:4} \eeq where the values of the parameters
are given in Table \ref{appendixa:tab:1}

\begin{table}
\begin{center}
\begin{tabular}{|c||c|c|c|} \hline
\hspace{1.5cm}  & \hspace{.3cm}  $f_+$ \hspace{.3cm} &  $f_0$
& \hspace{.3cm} $f_T$ \hspace{.3cm}  \\  \hline \hline
$F(0)$ &  0.319  &  0.319   &  0.355   \\
$c_1$ &  1.465  &  0.633   &  1.478   \\
$c_2$ &  0.372  & - 0.095  &  0.373   \\
$c_3$ &  0.782  &   0.591  &  0.700   \\ \hline
\end{tabular}
\caption{Form factors for $B \to K$ transition}
\label{appendixa:tab:1}
\end{center}
\end{table}

\section{\label{appendix:b} Input parameters}

\begin{center}
$m_B ~=~ 5.26$GeV \ , \ $m_b ~=~ 4.8$GeV  \ , \
$ m_c ~=~ 1.4$GeV  \\
$m_\mu ~=~  0.106$GeV  \ , \ $m_\tau ~=~ 1.77$GeV  \ , \
$m_w ~=~ 80.4$GeV  \ , \ $m_z ~=~ 91.19$GeV   \\
$V_{tb} V^*_{ts} = 0.0385$ \ , \ $\alpha = {1 \over 129}$  \ , \
$m_{K} = 0.49$GeV \ , \  $\Gamma_B = 4.22 \times 10^{-13}$GeV  \\
$G_F = 1.17 \times 10^{-5} ~{\rm GeV}^{-2}$
\end{center}


\section*{Acknowledgments}
The authors would like to thank Frank Kr\"{u}ger for his useful
remarks on the first version of the manuscript. The work of SRC
and NG was supported under the SERC scheme of the Department of
Science \& Technology (DST), India.  SRC wishes to thank KIAS,
Republic of Korea, for their hospitality during his visit there,
where this work was initiated. This work is partially supported by
Australian Research Council and grant from the University of
Melbourne.


\pagebreak


\bfig \centering{ \epsfig{file=drate_s.eps,height=.35\textheight}
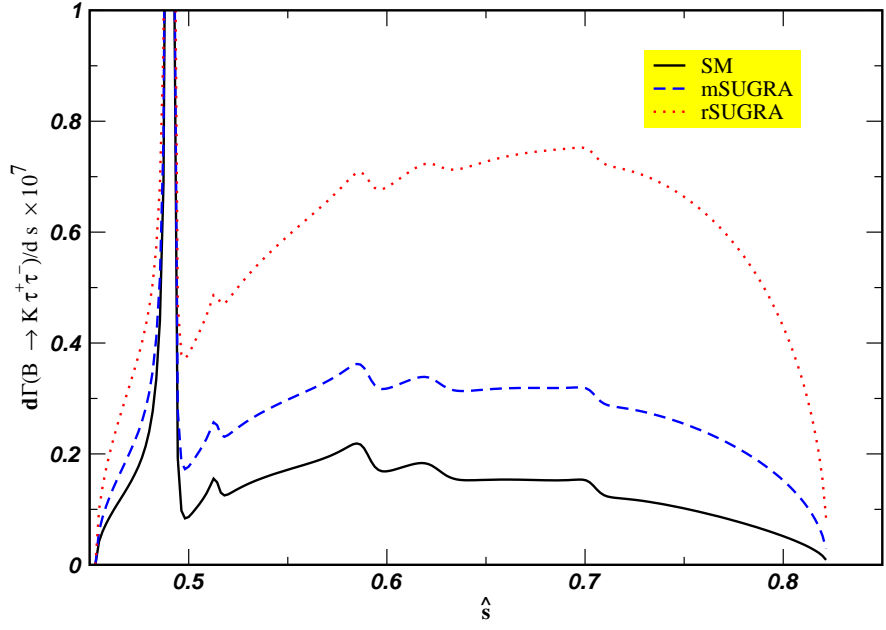
\caption{The differential decay rate $d\Gamma/d\sh$ against $\sh$.
The parameters for the mSUGRA model are: $m_0 = 400$GeV, $m_{1/2}
= 500$GeV, $tan\beta = 40$ and $A = 0$. The additional parameter
for the rSUGRA models is taken to be $m_A = 345$GeV.}
\label{fig:1}} \vskip 1cm \efig
\bfig \epsfig{file=drtb.eps,height=.35\textheight}
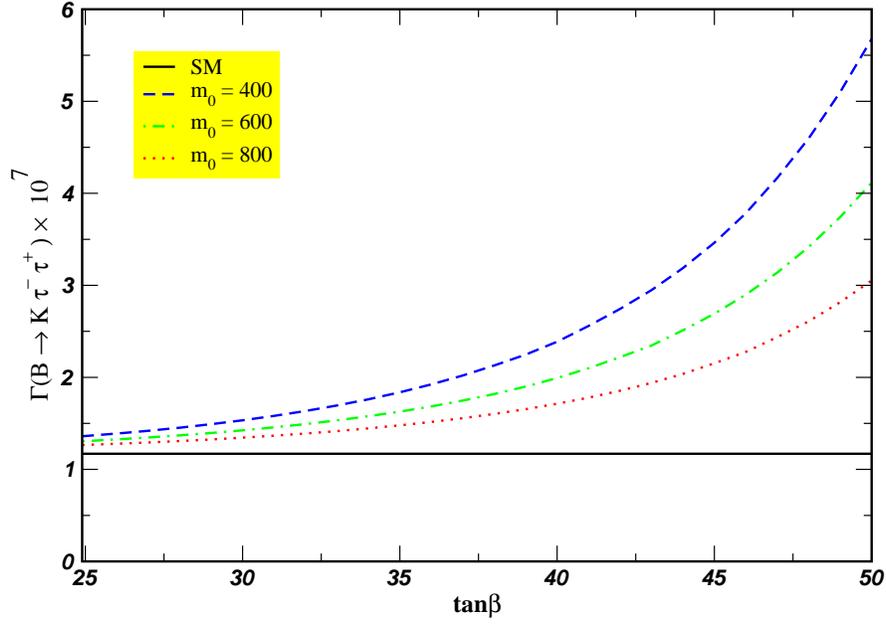
\caption{Br(\btoktt) variation with $tan\beta$ in mSUGRA for
various values of $m_0$. Other model parameters are: $m_{1/2} =
500$GeV, $A =0$.} \label{fig:2} \efig
\bfig \epsfig{file=drma.eps,height=.35\textheight}
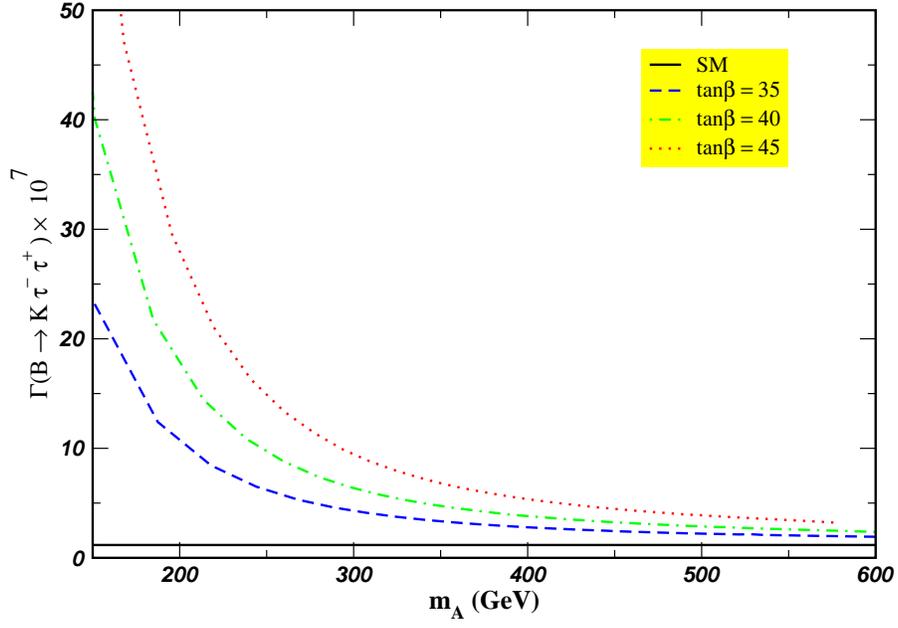
\caption{Br(\btoktt) variation with $m_A$ in rSUGRA for various
$tan\beta$ values. Other model parameters: $m_0 = 500$GeV,
$m_{1/2} = 500$ GeV, $A = 0$.} \label{fig:3} \efig
\begin{figure}
\epsfig{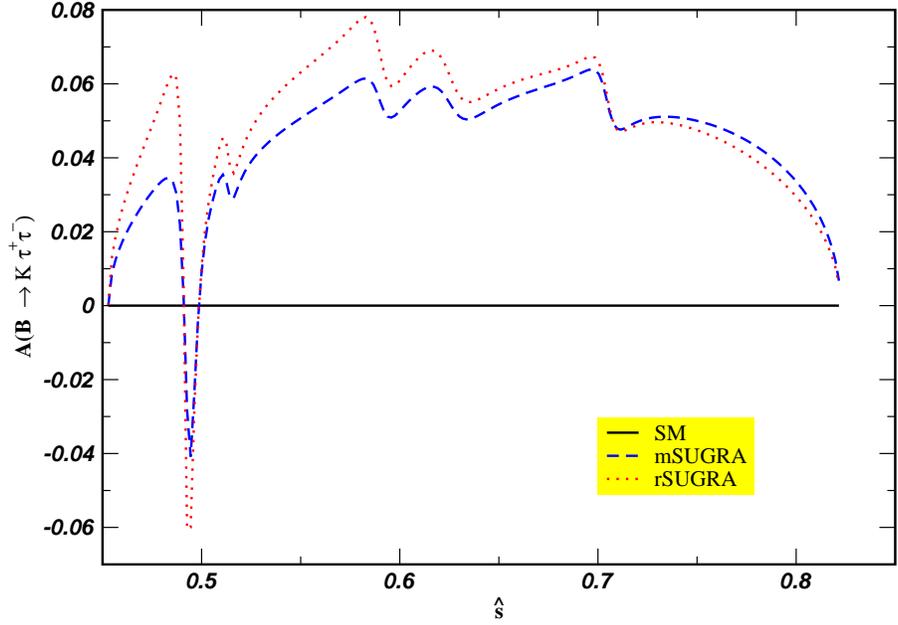} \caption{Unpolarized
FB asymmetry with dilepton invariant mass $\sh$. Other parameters
of mSUGRA and rSUGRA are given in figure \ref{fig:1}.}
\label{fig:4}
\end{figure}
\bfig \epsfig{file=fbtb.eps,height=.35\textheight}
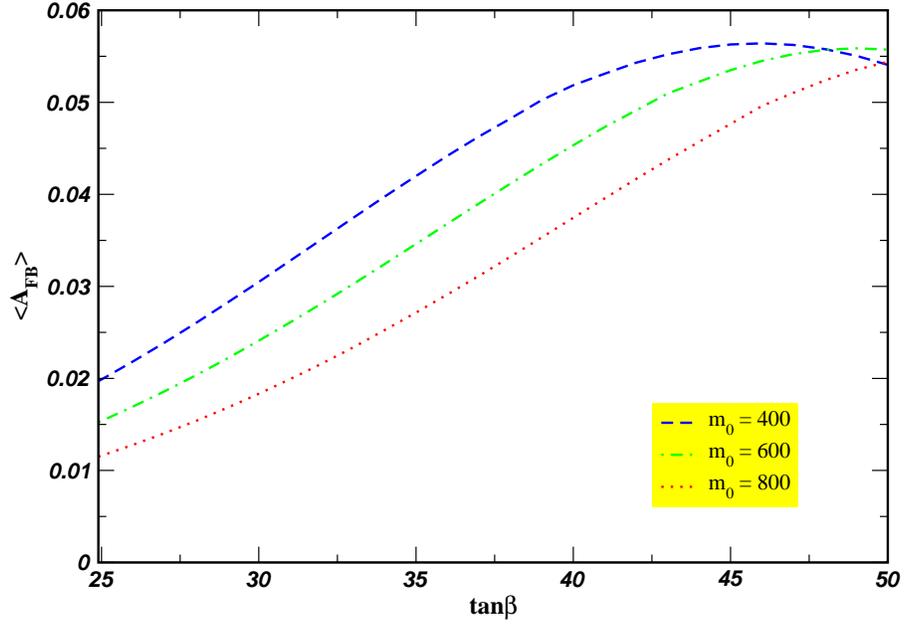
\caption{Integrated FB asymmetry variation with $tan\beta$ in
mSUGRA model. Other parameters are same as given in figure
\ref{fig:2}.} \label{fig:5} \efig
\bfig \epsfig{file=fbma.eps,height=.35\textheight}
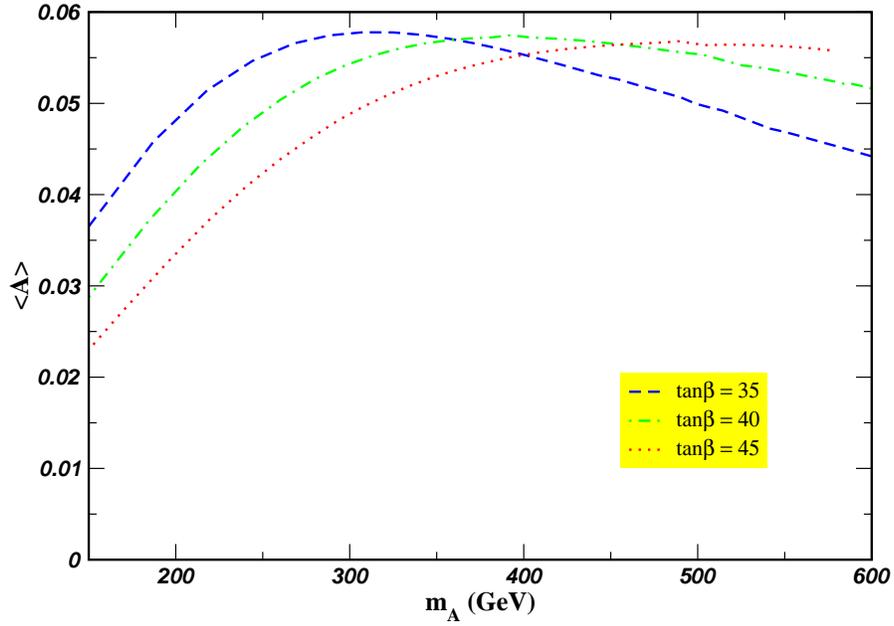
\caption{Integrated FB asymmetry variation with $m_A$ in rSUGRA
model, with other model parameters as given in figure
\ref{fig:3}.} \label{fig:6} \efig
\bfig \epsfig{file=fb_lmlong.eps,height=.35\textheight}
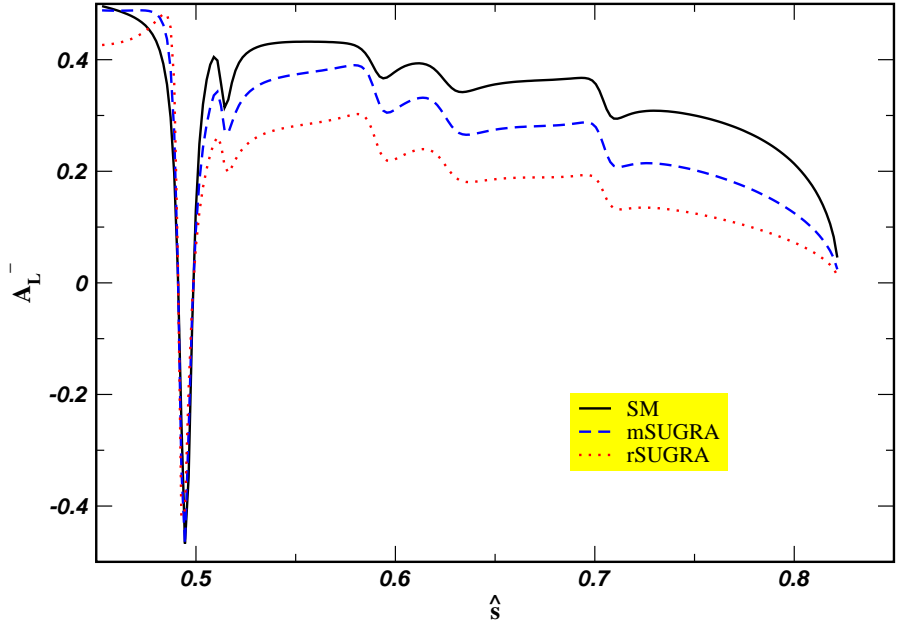
\caption{$A_L^-$ variation with $\sh$. Other parameters of mSUGRA
and rSUGRA model are the same as given in figure \ref{fig:1}.}
\label{fig:7} \efig
\bfig \epsfig{file=fb_lmlongtb.eps,height=.35\textheight}
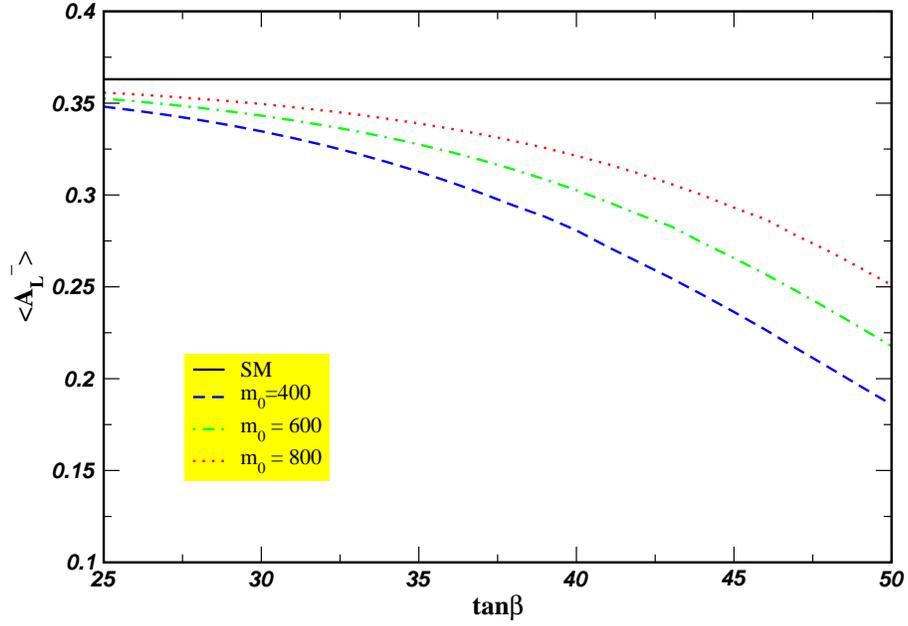
\caption{$<A_L^->$ variation with $tan\beta$ in mSUGRA model for
various values of $m_0$. Other model parameters are as given in
figure \ref{fig:2}.} \label{fig:8} \efig
\bfig \epsfig{file=fb_lmlongma.eps,height=.35\textheight}
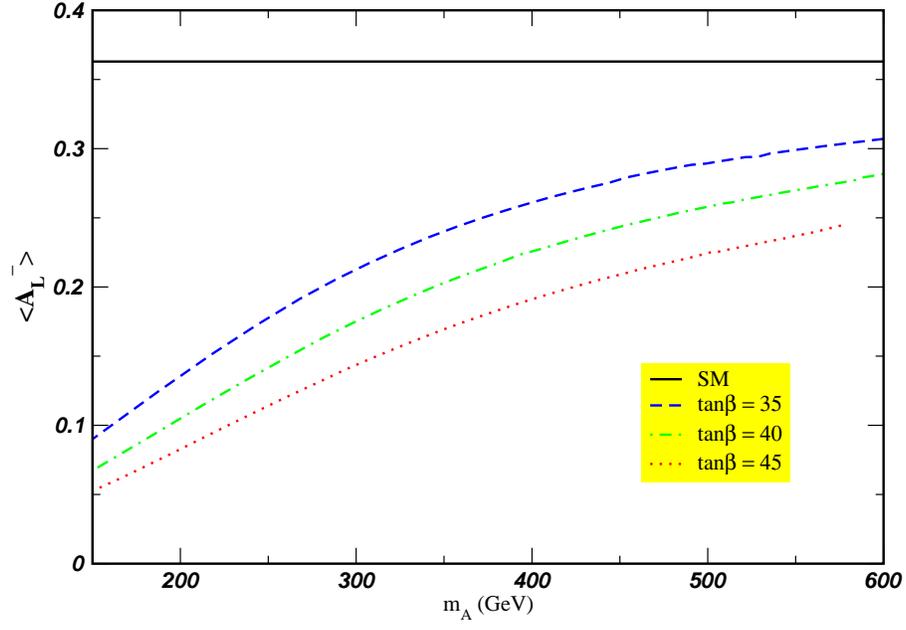
\caption{$<A_L^->$ variation with $m_A$ in rSUGRA model for
various values of $tan\beta$. Other model parameters are as given
in figure \ref{fig:3}.} \label{fig:9} \efig
\begin{figure}
\epsfig{file=fb_lmtrans.eps,height=.35\textheight}
\caption{$A_T^-$ with $\sh$. Other parameters are the same as
given in figure\ref{fig:1}.} \label{fig:10}
\end{figure}
\clearpage \bfig
\epsfig{file=fb_lmtranstb.eps,height=.35\textheight}
\caption{$<A_T^->$ with $tan\beta$ in mSUGRA, with other
parameters same as in figure \ref{fig:9}. } \label{fig:11} \vskip
1cm \efig
\bfig \epsfig{file=fb_lmtransma.eps,height=.35\textheight}
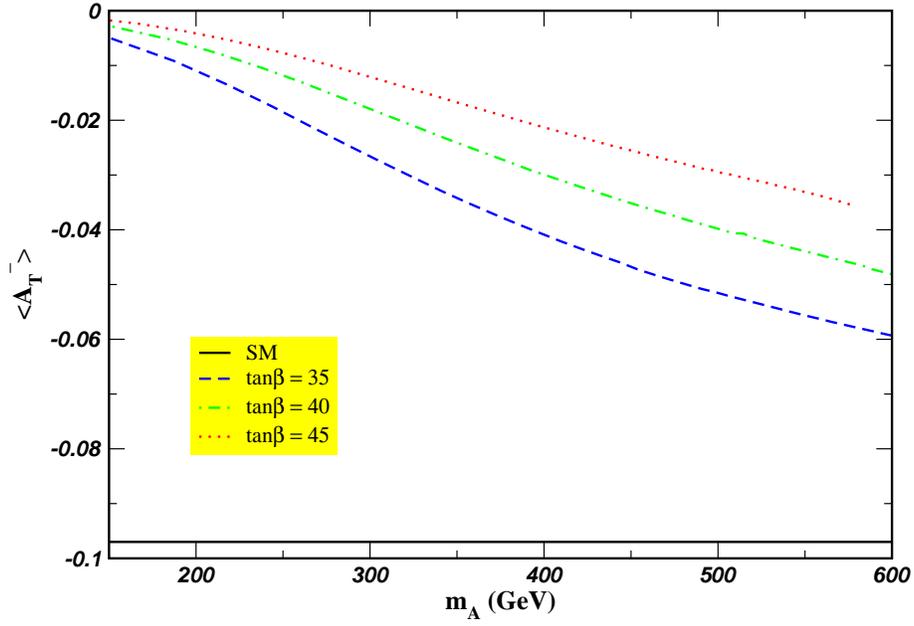
\caption{$<A_T^->$ with $m_A$ in rSUGRA, with other parameters
same as in figure \ref{fig:10}.} \label{fig:12} \efig
\begin{figure}
\epsfig{file=fb_lmlong_lpnorm.eps,height=.35\textheight}
\caption{$A_{LN}$ with $\sh$. Other parameters are the same as
given in figure \ref{fig:1}.} \label{fig:13}
\end{figure}
\bfig \epsfig{file=fb_lmlong_lpnormtb.eps,height=.35\textheight}
\caption{$<A_{LN}>$ with $tan\beta$ in mSUGRA, with other
parameters Same as in figure \ref{fig:9}.} \label{fig:14} \efig
\bfig \epsfig{file=fb_lmlong_lpnormma.eps,height=.35\textheight}
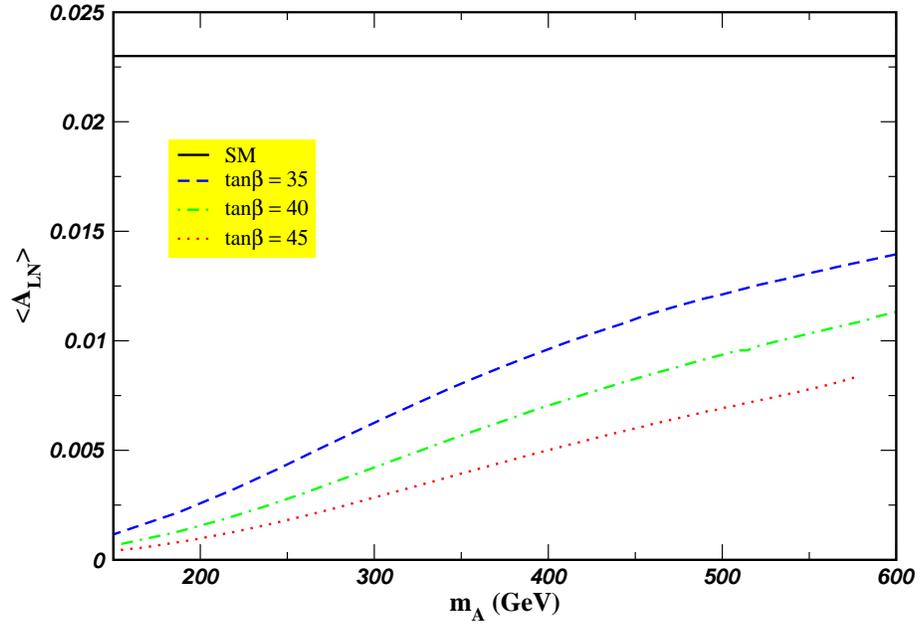
\caption{$<A_{LN}>$ with $m_A$ in rSUGRA model, with other
parameters same as in figure \ref{fig:10}.} \label{fig:15} \efig
\bfig \epsfig{file=fb_lmlong_lptrans.eps,height=.35\textheight}
\caption{$A_{LT}$ with $\sh$. Other parameters are the same as
given in figure \ref{fig:1}.} \label{fig:16}
\end{figure}
\bfig \epsfig{file=fb_lmlong_lptranstb.eps,height=.35\textheight}
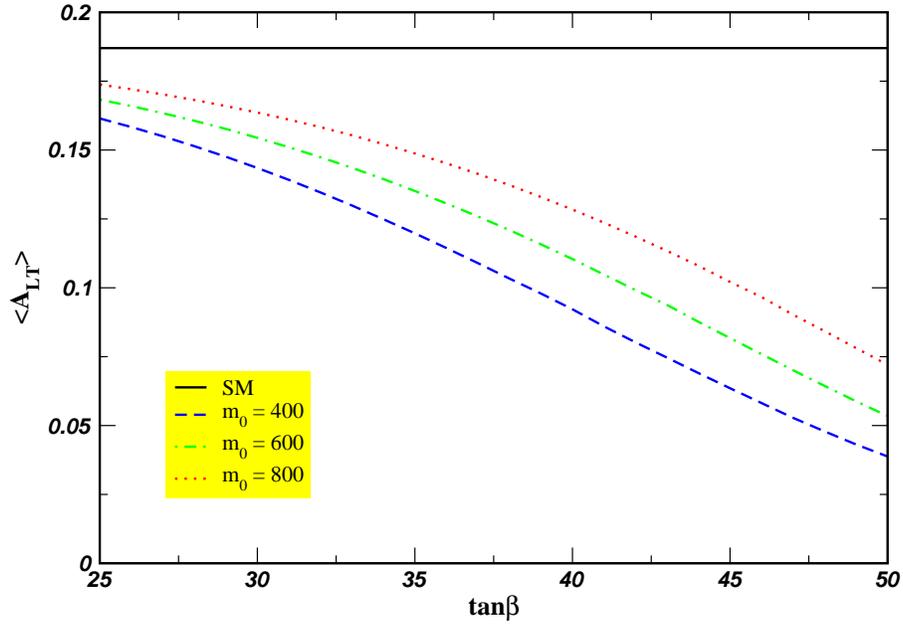
\caption{$<A_{LT}>$ with $tan\beta$ in mSUGRA, with the other
parameters being the same as in figure \ref{fig:9}.}
\label{fig:17} \efig
\bfig \epsfig{file=fb_lmlong_lptransma.eps,height=.35\textheight}
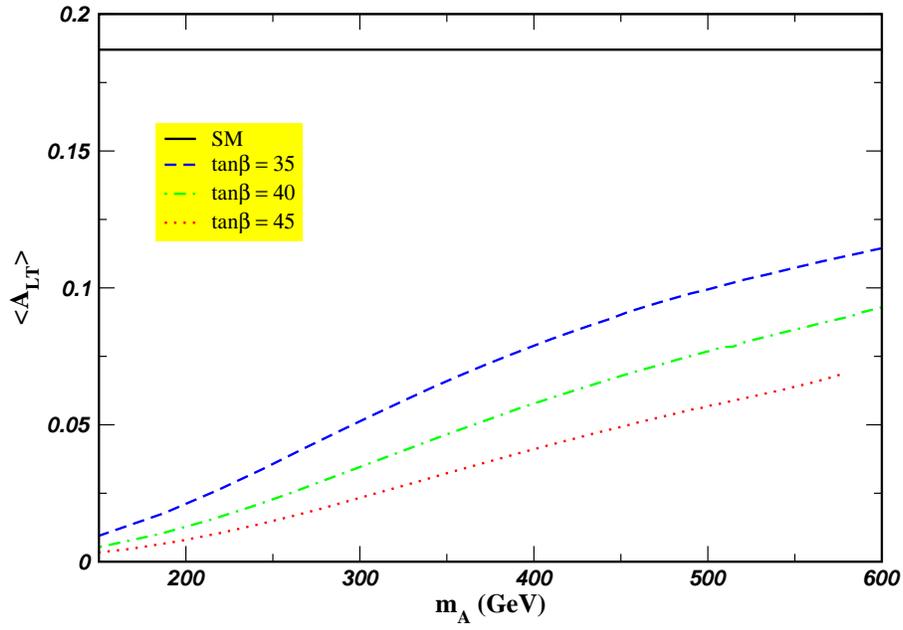
\caption{$<A_{LT}>$ with $m_A$, with the other parameters the same
as in figure \ref{fig:10}.} \label{fig:18} \efig
\clearpage \bfig
\epsfig{file=fb_lmnorm_lptrans.eps,height=.35\textheight}
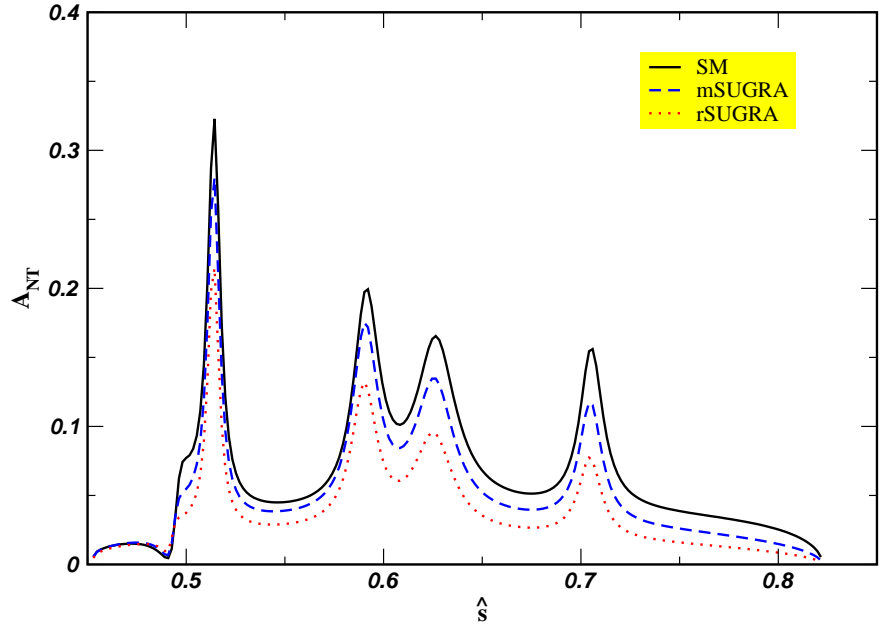
\caption{$A_{NT}$ with $\sh$. Other parameters are the same as
given in figure \ref{fig:1}.} \label{fig:19} \vskip 1cm \efig
\bfig \epsfig{file=fb_lmnorm_lptranstb.eps,height=.35\textheight}
\caption{$<A_{NT}>$ with $tan\beta$ in mSUGRA, with other
parameters same as in figure \ref{fig:9}.} \label{fig:20} \efig
\bfig \epsfig{file=fb_lmnorm_lptransma.eps,height=.35\textheight}
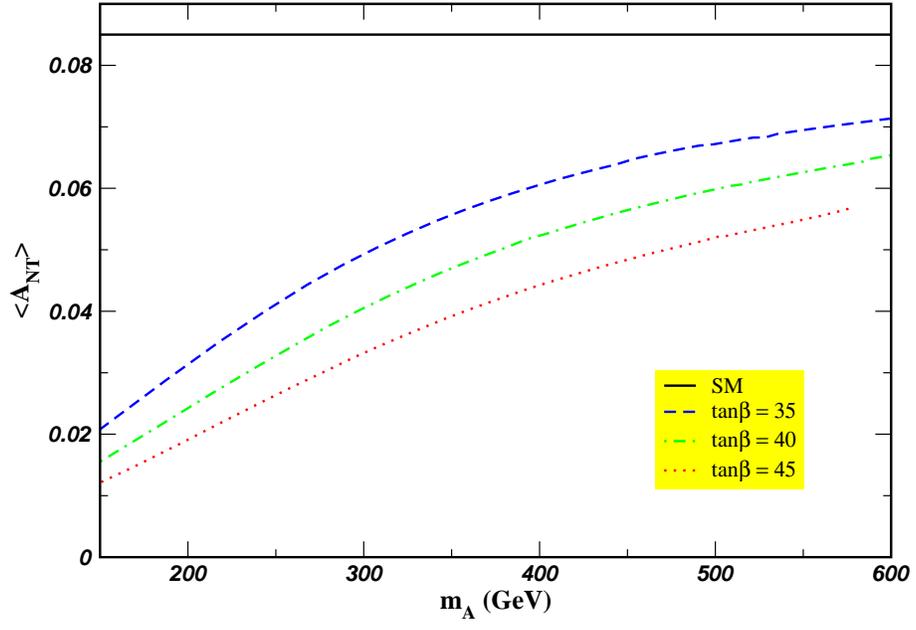
\caption{$<A_{NT}>$ with $m_A$ in rSUGRA model, with the other
parameters the same as in figure \ref{fig:10}.} \label{fig:21}
\efig

\end{document}